# Orthogonal Extended Infomax Algorithm

Nicole Ille

BESA GmbH, Gräfelfing, Germany
nille@besa.de

**Abstract**

The extended infomax algorithm for independent component analysis (ICA) can separate sub- and super-Gaussian signals but converges slowly as it uses stochastic gradient optimization. In this paper, an improved extended infomax algorithm is presented that converges much faster. Accelerated convergence is achieved by replacing the natural gradient learning rule of extended infomax by a fully-multiplicative orthogonal-group based update scheme of the unmixing matrix leading to an orthogonal extended infomax algorithm (OgExtInf). Computational performance of OgExtInf is compared with two fast ICA algorithms: the popular FastICA and Picard, a L-BFGS algorithm belonging to the family of quasi-Newton methods. Our results demonstrate superior performance of the proposed method on small-size EEG data sets as used for example in online EEG processing systems, such as brain-computer interfaces or clinical systems for spike and seizure detection.

# 1. Introduction

Independent component analysis (ICA) attempts to solve the blind source separation (BSS) problem by decomposing a mixture signal into maximally statistically independent source signals using higher-order statistics [1] [2].

ICA is widely used in multivariate data analysis: for example in EEG and MEG analysis, to explore data or to correct artifacts [3] [4] [5] [6], in functional magnetic resonance imaging (fMRI) [7][8], in financial time series analysis [9][10], in biology [11][12], or analytical chemistry [13][14] [15]. It has also been used in brain-computer interfaces (BCI) to remove artifacts or to enhance task-related EEG signals [16].

Popular linear ICA algorithms are JADE [17], FastICA [18] [19], infomax [20], and extended infomax [21]. JADE requires explicit calculation of fourth order statistics. In FastICA, infomax, and extended infomax, higher-order statistics are incorporated by non-linear functions derived from the chosen probability density model. Infomax uses only one non-linearity, which is derived from a super-Gaussian probability density function (PDF). Extended infomax, on the other hand, can separate mixtures of sub- and super-Gaussian signals by switching between two non-linear functions. FastICA can also deal with sub- and super-Gaussian components. This is useful, e.g., to remove artifacts with sub-Gaussian PDF like powerline artifacts from EEG data.

While FastICA is a fixed-point algorithm with very fast convergence especially on data which is a mixture of truly independent components [22] [23], infomax and extended infomax converge relatively slowly as they are based on stochastic gradient optimization [24].

To accelerate convergence of gradient descent, quasi-Newton methods may be used, like for example the 'preconditioned ICA for real data' (Picard) algorithm [25] or Picard-O under orthogonal constraint [26]. These apply the quasi-Newton L-BFGS solver preconditioned with Hessian approximations and are also available in extended versions to separate sub- and super-Gaussian signals. On real data, where the independence assumption may not hold perfectly, it has been shown that Picard-O is more consistent in its convergence pattern and can be much faster than FastICA [26].

In a previous work [27], an alternative fully-multiplicative orthogonal-group ICA neural algorithm with fast convergence was introduced. However, this algorithm and a related method [28] can only separate super-Gaussian signals. In the FastAdaptiveOgICA algorithm [29], the fully-multiplicative orthogonal-group approach was extended to separate sub-Gaussian, super-Gaussian, and near-Gaussian source signals by switching between three different non-linear functions. The non-linearities and the switching rule used in this approach are impractical, however, as they are based on unhandy lookup tables and sample moments and may lead to instabilities of the separating process.

In this paper, we propose to combine the fully-multiplicative orthogonal-group ICA neural algorithm [27] with the non-linear functions of extended infomax (ExtInf) to achieve fast and stable separation of sub- and super-Gaussian signals. The resulting orthogonal extended infomax algorithm (OgExtInf) is briefly described below, and differences to previous related work are outlined. Moreover, a derivation of the fully-multiplicative update scheme is given. The proposed method is then applied to simulated data and to real EEG data from clinical and BCI contexts. Finally, the computational performance of OgExtInf is compared with reference methods ExtInf, FastICA, and the extended versions of Picard and Picard-O.

## 2. Methods

The ICA of a data matrix $D = (d_1, \ldots, d_n)^T \in R^{n \times t}$ of $n$ channels and $t$ time points is obtained by finding a linear transformation matrix $W \in R^{n \times n}$ such that the output source signals $S = (s_1, \ldots, s_n)^T \in R^{n \times t}$ determined by

$$S = WD \quad (1)$$

are mutually statistically independent. The matrix $W$ is called separation or unmixing matrix.

In the ExtInf algorithm [21], the unmixing matrix is updated in each iteration starting from an initial guess according to the conventional gradient learning rule

$$W = W + eps * \Delta W$$

where *eps* is the learning rate. The additive update $\Delta W$ is determined by following the natural gradient, i.e., the steepest direction of the cost function [30]

$$\Delta W \propto [I - \varphi(s)s^T]W$$

where $\varphi(s) = [\varphi(s_1), \ldots, \varphi(s_n)]^T$ is a component- and sample-wise non-linear function derived from the chosen PDF. ExtInf switches for each source signal $s_i, i = 1 \ldots n$ between a pair of non-linear functions $\varphi(s)$ derived from a super- and a sub-Gaussian PDF. The PDFs are selected such that the non-linear functions differ only in sign parameter $k$

$$\varphi(s) = s + k \tanh(s) \quad \begin{cases} k = +1: \ super - Gaussian \\ k = -1: \ sub - Gaussian \end{cases} \quad (2)$$

The data-driven switching rule of ExtInf

$$k = sign(E\{\text{sech}^2(s)\}E\{s^2\} - E\{\tanh(s)\,s\}) \qquad (3)$$

is based on stability analysis [31].

Under the orthogonality or whiteness constraint, the second-order covariance matrix R of the source signals is equal to the identity matrix

$$R = E\{ss^T\} = I$$

From the Bussgang property for blind deconvolution [32]

$$E\{s_i s_j^T\} = E\{f(s_i)s_j^T\}$$

with non-linear function $f(.)$ and independent stochastic processes $i, j$, it is obvious that under the whiteness constraint the same relationship holds for the non-linearized covariance matrix, which incorporates higher-order cross-correlations between the source signals due to the non-linear transformation:

$$\hat{R} = \begin{bmatrix} E\{f(s_1)s_1^T\} & & 0 \\ & \ddots & \\ 0 & & E\{f(s_n)s_n^T\} \end{bmatrix} = I$$

From $R = \hat{R} = I$ we obtain (cf. Appendix)

$$W^{-1}\hat{R} = W^{-1}$$

which can be rewritten as

$$W = \hat{R}^{-1}W$$

leading finally to the fully-multiplicative iterative update scheme suggested in [27] and [28] that successively diagonalizes the higher-order covariance matrix:

$$\widetilde{W}_{n+1} = \hat{R}_n^{-1} W_n$$

$$W_{n+1} = \widetilde{W}_{n+1}\left(\widetilde{W}_{n+1}^T \widetilde{W}_{n+1}\right)^{-1/2} \qquad (4)$$

The last expression is a symmetric orthogonalization step that is necessary to project each partial update matrix back into the orthogonal group. Note that this algorithm needs prewhitening of the data. For ExtInf, prewhitening is not a prerequisite but improves convergence.

In previous work [27], two non-linear functions $f(s) = \tanh(2s)$ and $g(s) = s^2 \, sign(s)$ were used to build the higher-order covariance matrix $E\{f(s)g(s^T)\}$. In another work [28], non-linear function $f(s) = \tanh(s)$ and linear identity function $g(s) = s$ are suggested. Note that using $g(s) = s$ is equivalent to the above formulation of $\hat{R}$. However, with the non-linear functions of [27] and [28] only super-Gaussian signals can be separated.

We propose to combine the fully-multiplicative iterative update scheme of equation (4) with non-linear function $f(s) = \varphi(s)$ of ExtInf given in equation (2). Diagonalization of $\hat{R}$ is then equivalent to the ExtInf cost function. For less than 1000 samples, we apply the switching rule of ExtInf according to equation (3), otherwise the sign of sample excess kurtosis is used as a switching criterion [33]. The OgExtInf algorithm is summarized in Fig. 1.

---

**Input:** Mixed signals $D$, initial unmixing matrix $W_0$, number of iterations $K$
**for** $k=0,1,…, K-1$ **do**
    Set $S_k = W_k D$
    Compute non-linear transformation $\hat{S}_k$ of $S_k$ using eq. (2)
    Compute $\hat{R}_k = \hat{S}_k S_k^T$
    Compute inverse $\hat{R}_k^{-1}$ of $\hat{R}_k$
    Set $W_{k+1}$ using eqs. (4)
**end**
**Output:** $W_{k+1}$

---

Figure 1: The orthogonal extended infomax (OgExtInf) algorithm.

## 3. Results

*Test setup*

The proposed OgExtInf method was applied to simulated and real EEG data, and computational performance was compared with ExtInf, FastICA, and the extended versions of Picard and Picard-O.

In this study, the OgExtInf and ExtInf implementation of the Besa ICA module (Besa GmbH, Gräfelfing, Germany) was used. Computations are performed in double precision and numerical code is optimized by Intel MKL 2020 with OpenMP threading in this module. For comparability, ExtInf and OgExtInf use the same accelerations and call the same functions. Weight change and computation time were recorded for every iteration to measure performance.

In addition, the parallel symmetric FastCA implementation of Sklearn and the Picard implementation of https://github.com/pierreablin/picard were used. Only the total computation time over all iterations could be recorded here.

All tests were run on a standard computer with an Intel Core i7-7700 CPU @ 3.6 GHz with 4 cores.

*Simulated data*

Two sets of synthetic EEG data were created. For experiment 1, 100 synthetic data matrices with 20 channels and 5000 random samples were calculated from 20 independent sources by multiplying the source signals with a random 20 x 20 mixing matrix whose entries are normally distributed with zero mean and unit variance. 10 sources have super-Gaussian Laplacian density PDF $p(x) = 0.5\, e^{-|x|}$, and 10 sources have sub-Gaussian uniform density PDF $p(x) = 0.25$ for x ≥ -2 and x ≤ 2.

For experiment 2, 100 synthetic data matrices with 50 channels and 10000 random samples were created accordingly from 50 independent sources with 25 sources having super-Gaussian Laplacian density and 25 sources having sub-Gaussian uniform density as in experiment 1.

Data was centered and whitened. Dimensionality of data was not reduced in the simulation study. The algorithms stopped if weight change was ≤ 1.e$^{-6}$ or after 1000 iterations.

Computation time and weight change of OgExtInf and ExtInf versus number of iterations are shown in Fig. 2 across all simulated files. While OgExtInf converges for 50% of the simulated data sets within 187 iterations (67 ms) in experiment 1 and 356 iterations (770 ms) in experiment 2, ExtInf does not reach the convergence criterion within 1000 iterations corresponding to 4.2 s in experiment 1 and 44 s in experiment 2.

In Fig. 3, the time until convergence of OgExtInf is compared with Picard, Picard-O, and FastICA. Here, FastICA performs best for the simulated data in both experiments with a median run time of 31 ms and 218 ms. This is in accordance with previous reports where FastICA converged very fast on data which is an actual mixture of independent components [22] [23]. For the smaller data set in experiment 1 with 5000 samples, OgExtInf has second fastest convergence with a median run time of 67 ms.

Finally, we investigated how well the algorithms recover the simulated mixture of independent sources. To this end, we calculated the Amari distance [34] between the known truth mixing matrix $A$ and the estimated unmixing matrix $W$ according to [35]:

$$\frac{1}{2n}\sum_{i=1}^{n}\left(\frac{\sum_{j=1}^{n}|R_{ij}|}{max_j|R_{ij}|}-1\right) + \frac{1}{2n}\sum_{j=1}^{n}\left(\frac{\sum_{i=1}^{n}|R_{ij}|}{max_i|R_{ij}|}-1\right)$$

with $R = WA$. This metric is invariant to permutation and scaling of the columns of the mixing and unmixing matrix. Moreover, it is always between 0 and $(n -$

1) and is equal to zero only if mixing and unmixing matrix represent the same components [35]. From Fig. 4 it can be seen that OgExtInf recovers the simulated sources equally well as Picard, Picard-O, and FastICA, while the Amari distance of ExtInf, which did not converge within 1000 iterations, is considerably higher in both simulated experiments.

*Real clinical EEG data*

Next, the algorithms were run on 12 real EEG data segments extracted from 12 clinical EEGs of the open-source Temple University Hospital EEG Corpus [36]. Each extracted data segment consists of 23 recording channels (according to the 10-20 system) with a duration of 40 s at a sampling rate of 250 Hz and is high-pass filtered at 0.5 Hz. The data sets contain background EEG and artifacts like EOG, muscle, powerline, and ECG. Four of the data sets also include spikes. The ICA algorithms were run on the entire 40 s (corresponding to 10000 samples), the first 30 s (7500 samples), the first 20 s (5000 samples), and the first 10 s (2500 samples) of the extracted data segments.

Prior to ICA, data was again centered and whitened. This time, dimensionality of data was reduced by PCA retaining all PCA components explaining at least 1% of the variance in the input data, resulting in 11 to 21 PCA components for the current data. The ICA algorithms stopped if weight change was $\leq 1.e^{-6}$ or after 3000 iterations.

In Fig. 5, the median time until convergence for OgExtInf, Picard, Picard-O, and FastICA is displayed. The proposed ExtOgInf algorithm converges faster or equally fast as the reference methods. For 2500 samples, OgExtInf converges 2.6 times faster (within 109 ms) than the second fastest algorithm Picard-O (285 ms). As FastICA did not converge for 6 data sets with 2500 samples, the median convergence time of FastICA is relatively high in this case. FastICA did also not converge for 2 data sets with 5000 samples and one data set with 7500 samples and 10000 samples in this study. This is in accordance with previous studies reporting varying convergence patterns of FastICA for real data [26].

*Real BCI data*

Finally, the algorithms were run on continuous BCI data from the BCI Competition IV. We used the calibration data of data set 1, which was recorded during a cued motor imagery (MI) task across 7 subjects [37]. Cues were displayed for a period of 4 s during which the subject was instructed to imagine a movement of the hand (left or right) or the foot. These periods were interleaved with 2 s of blank screen and 2 s with a fixation cross shown in the centre of the screen. EEG signals were recorded with 59 electrodes at a sampling rate of 1000 Hz. For our experiment, we extracted 10 s of data including imagined movement

(seconds 17 to 27). The ICA algorithms were run on the 10 s segment at 1000 Hz, and down-sampled to 750 Hz, 500 Hz, and 250 Hz. Before down-sampling data was low-pass filtered with a high-cutoff set to 2/3 of the new sampling frequency. Additionally, all data sets are high-pass filtered at 0.5 Hz.

Prior to ICA, data was again centered and whitened. Moreover, dimensionality of data was reduced by PCA retaining all PCA components explaining at least 1% of the variance in the input data, resulting in 11 to 23 PCA components for the current data. The ICA algorithms stopped if weight change was $\leq 1.e^{-6}$ or after 3000 iterations.

In Fig. 6, the median time until convergence for OgExtInf, Picard, Picard-O, and FastICA is displayed. In all cases, the proposed ExtOgInf algorithm converges faster than the other methods. For 2500 samples, OgExtInf converges about 2 times faster (within 134 ms) than the second fastest algorithm FastICA (265 ms). Again, FastICA did not converge for all data sets (2 for 2500, 3 for 5000, 2 for 7500, and 3 for 10000 samples).

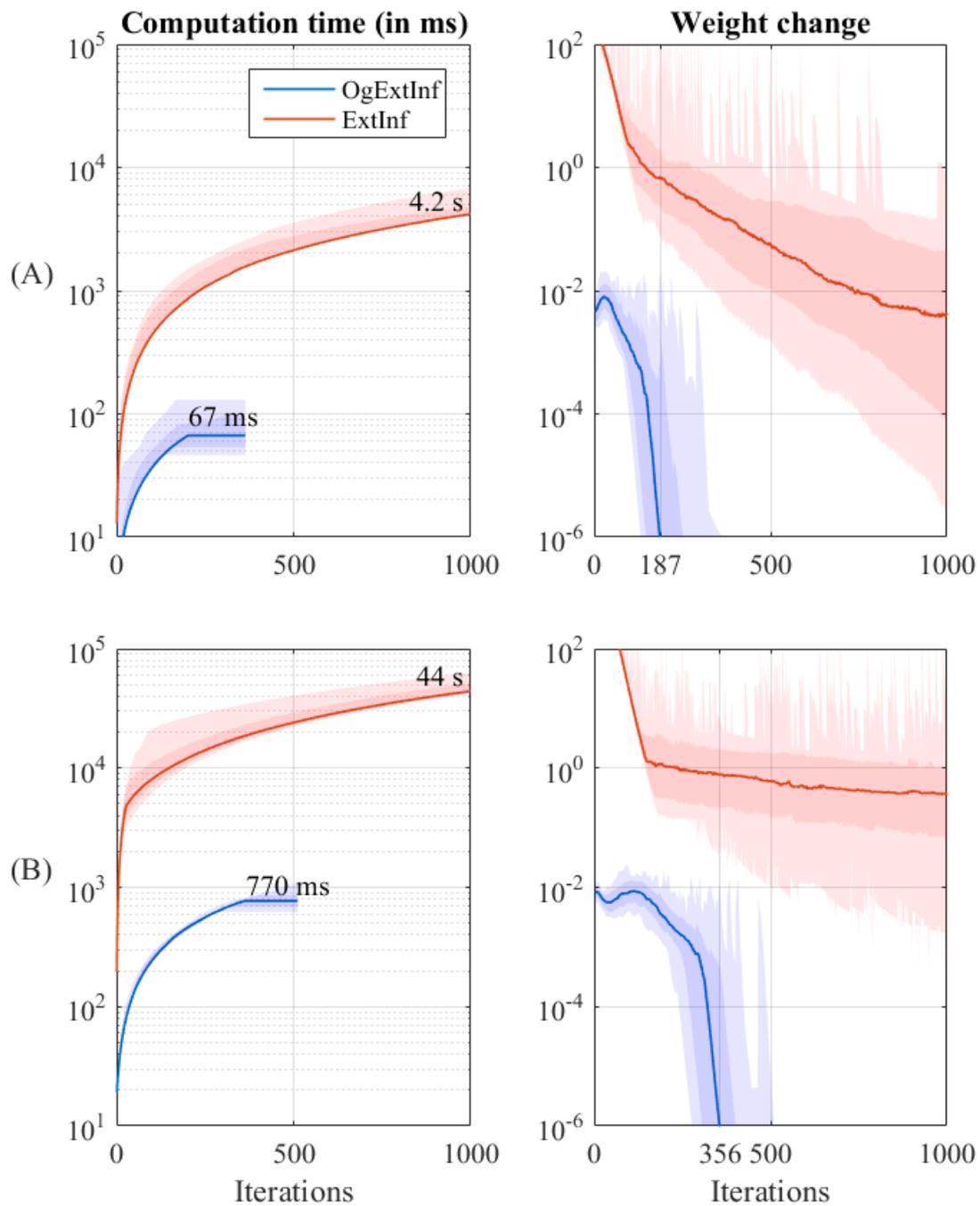

Figure 2: Comparison of computation time and weight change versus number of iterations for OgExtInf (blue) and ExtInf (red) on simulated data in (A) experiment 1 with 20 simulated sources and (B) experiment 2 with 50 simulated sources. Solid lines and numbers correspond to the median of the 100 runs, the darker shaded areas cover the 10% to 90% percentiles, the lighter shaded areas range from minimum to maximum.

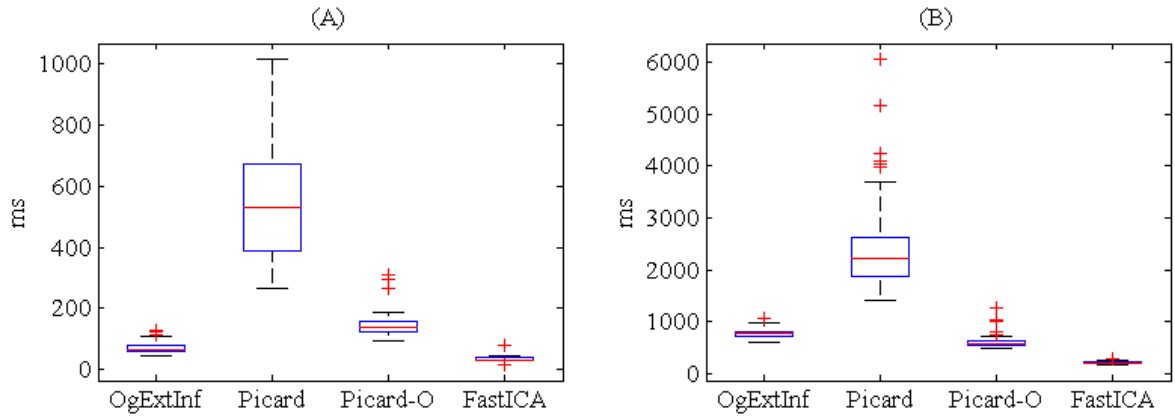

Figure 3: Box plots of time until convergence in milliseconds for OgExtInf, Picard, Picard-O, and FastICA on 100 simulated data sets in (A) experiment 1 with 20 sources and 5000 samples and (B) experiment 2 with 50 sources and 10000 samples. The median times (from left to right) in experiment (A) are 67 ms, 531 ms, 140 ms, 31 ms and in experiment (B) are 770 ms, 2.23 s, 593 ms, and 218 ms.

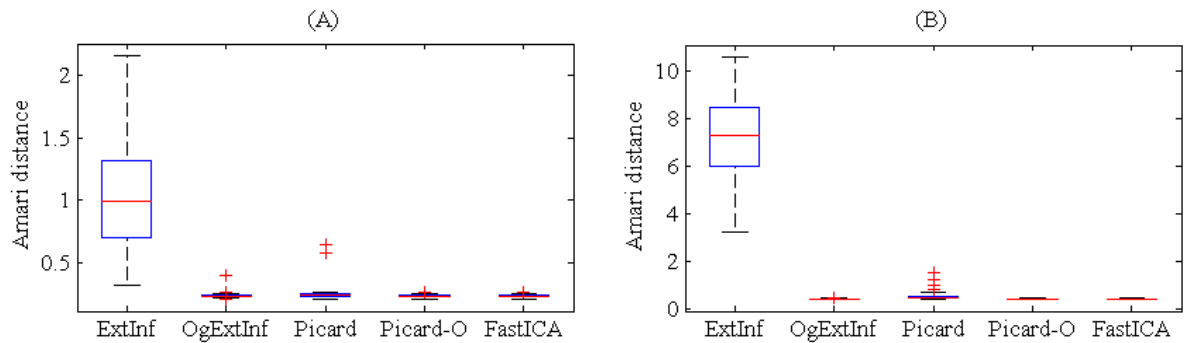

Figure 4: Amari distance for ExtInf, OgExtInf, Picard, Picard-O, and FastICA on 100 simulated data sets in (A) experiment 1 with 20 sources and 5000 samples and (B) experiment 2 with 50 sources and 10000 samples.

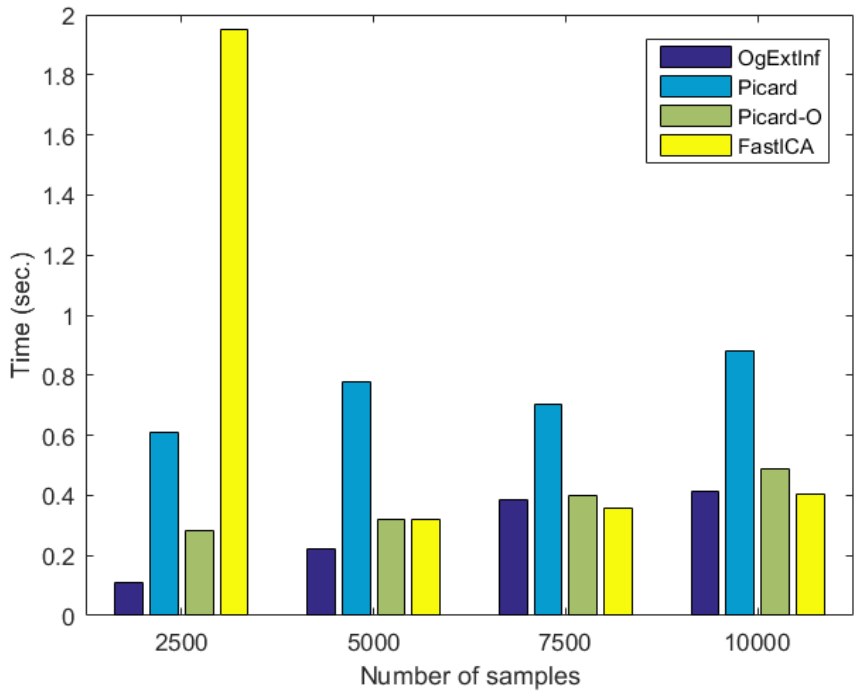

Figure 5: Median time in seconds until convergence criterion is reached by OgExtInf, Picard, Picard-O, and FastICA for 12 real clinical EEG data sets with 23 channels and 2500, 5000, 7500, and 10000 samples using 11 to 21 components as determined by 1% PCA criterion.

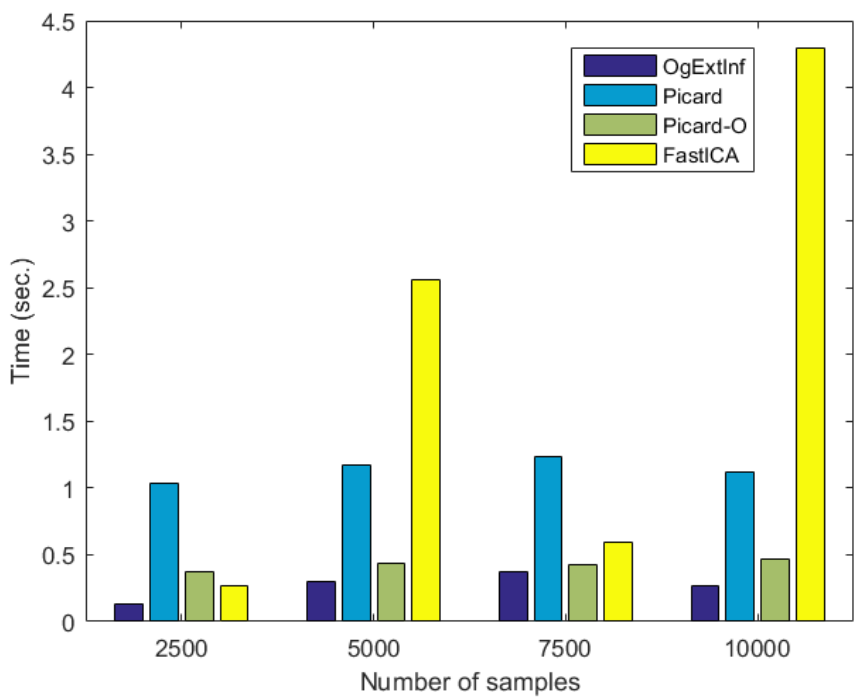

Figure 6: Median time in seconds until convergence criterion is reached by OgExtInf, Picard, Picard-O, and FastICA for 7 real BCI data sets with 59 channels and 2500, 5000, 7500, and 10000 samples using 11 to 23 components as determined by 1% PCA criterion.

## 4. Discussion

In this paper, the OgExtInf algorithm was presented, which minimizes the extended infomax cost function over the space of orthogonal weight matrices using the fully-multiplicative orthogonal-group update scheme of the weight matrix proposed in [27]. This modification improves convergence behavior and convergence speed of the original extended infomax algorithm considerably.

Our numerical comparison of OgExtInf with fast state-of-the-art ICA algorithms, FastICA, Picard-O, and Picard, on clinical EEG data and BCI data reveals comparable or superior convergence speed of OgExtInf. For small sample sizes, OgExtInf may converge about 2 to 3 times faster than the second fastest algorithm.

Our results also support what has been previously reported that FastICA has optimal convergence on simulated data where the independence assumption holds perfectly but may have impaired convergence for real data where this assumption is not completely fulfilled [26].

Comparative studies of ICA algorithms are often based on large-size data with more than 50 components and considerably more than 10000 samples occurring for example in high-density EEG recordings, functional MRI, or image data [25] [26]. In this work, we focus on online applications that process leaner EEG data such as practical BCI or clinical systems for spike and seizure detection.

In these applications, artifacts are often a serious problem, and artifact removal is required to improve performance or in BCI, to prevent that artifacts, for example muscle activity in motor imagery tasks, control the system [38]. ICA has proven to be a useful tool for removal of various types of EEG artifacts [39] [40] [41] but may not be readily applicable in online applications due to technical challenges [16]. Only few BCI studies implemented online ICA-based artifact removal [38] [42]. In these studies, ICA was calculated in short overlapping sliding windows of 1 s to 3 s. However, if the sliding window is too small, there may not be enough samples to reliably estimate the weights of the unmixing matrix [43] [44] [45]. With fast algorithms, ICA can be calculated on sufficiently long segments. In the BESA Artifact module (Besa GmbH, Gräfelfing, Germany), this approach has been elaborated for ongoing correction of EOG, ECG, muscle, and powerline artifacts using the proposed OgExtInf method [46].

Moreover, sufficiently fast ICA algorithms like OgExtInf may help to build adaptive ICA-based spatial filters or to extract ICA-based features from the ongoing EEG. These may be used as brain control signals in BCI systems or to detect spike and seizure activity during epilepsy monitoring.

## 5. Conclusion

In this paper, we propose the OgExtInf algorithm as a fast orthogonal variant with improved convergence of the well-known extended infomax algorithm. Experiments show that OgExtInf is reliable and faster as the widely used FastICA and Picard-O algorithms on small-size real EEG data sets. OgExtInf may therefore be useful to make ICA applicable in online applications like artifact correction, adaptive brain-computer interfaces, or epileptic spike and seizure detection.

## Appendix

Equation (1) can be extended to

$$SS^T = WDS^T$$

Substituting under the whiteness constraint

$$R = SS^T = I$$

we obtain

$$I = WDS^T$$

Setting $D = W^{-1}S$ and rearranging we get

$$W^{-1}R = W^{-1}$$

With the Bussgang property

$$R = E\{ss^T\} = E\{f(s)s^T\} = \hat{R}$$

this is equal to

$$W^{-1}\hat{R} = W^{-1}$$